\documentclass[aps,prl,twocolumn,balancelastpage,showpacs,amsfonts,amssymb,amsmath]{revtex4-1}

\usepackage[dvipdfmx]{graphicx}
\usepackage[dvipdfmx]{hyperref}

\newcommand{\NA}{\mathrm{NA}}
\newcommand{\Yb}{\mathrm{Yb}}
\DeclareMathSymbol{\mhyph}{\mathalpha}{operators}{`-}

\begin{document}

\onecolumngrid
\title{Site-resolved imaging of single atoms with a Faraday quantum gas microscope}
\author{Ryuta Yamamoto}
\altaffiliation{Electronic address: r\_yamamoto@scphys.kyoto-u.ac.jp}
\affiliation{Department of Physics, Graduate School of Science, Kyoto University, Kyoto 606-8502, Japan}
\author{Jun Kobayashi}
\affiliation{Department of Physics, Graduate School of Science, Kyoto University, Kyoto 606-8502, Japan}
\author{Kohei Kato}
\altaffiliation{Present address: Department of Physics, Osaka City University, Sumiyoshi-ku, Osaka 558-8585, Japan}
\affiliation{Department of Physics, Graduate School of Science, Kyoto University, Kyoto 606-8502, Japan}
\author{Takuma Kuno}
\affiliation{Department of Physics, Graduate School of Science, Kyoto University, Kyoto 606-8502, Japan}
\author{Yuto Sakura}
\affiliation{Department of Physics, Graduate School of Science, Kyoto University, Kyoto 606-8502, Japan}
\author{Yoshiro Takahashi}
\affiliation{Department of Physics, Graduate School of Science, Kyoto University, Kyoto 606-8502, Japan}
\date{\today}

\begin{abstract}
We successfully demonstrate a quantum gas microscopy using the Faraday effect which has an inherently non-destructive nature.
The observed Faraday rotation angle reaches 3.0(2) degrees for a single atom.
We reveal the non-destructive feature of this Faraday imaging method by comparing
the detuning dependence of the Faraday signal strength with that of the photon scattering rate.
We determine the atom distribution with deconvolution analysis.
We also demonstrate the absorption and the dark field Faraday imaging,
and reveal the different shapes of the point spread functions for these methods,
which are fully explained by theoretical analysis.
Our result is an important first step towards an ultimate quantum non-demolition site-resolved imaging
and furthermore opens up the possibilities for quantum feedback control of a quantum many-body system with a single-site resolution.
\end{abstract}
\maketitle

Measurement and manipulation of each single quantum object in a quantum many-body system lie at the heart of quantum information processing~\cite{NielsenChuang:QC2010}. 
For ultracold atoms in an optical lattice, a technique of single-site-resolved imaging and single-site-addressing,
called quantum gas microscope~(QGM), is recently demonstrated for bosons
\cite{Greiner:Rb-QGM2009, Bloch:Rb-QGM:SingleAddressing2011, Kozuma:Yb-QGM2015, RYamamoto:Yb-QGM2016}
and fermions
\cite{Zweirlein:K-QGM2015, Kuhr:K-QGM2015, Greiner:Li-QGM2015, Bloch:Li-QGM:PauliBlock2015, Thywissen:K-QGM2015}.
The development of QGM technique enables us to realize various fascinating experiments in the study of quantum many-body system
\cite{Ott:ReviewOfQGM2016, Greiner:Rb-QGM:Entaglement2015, Greiner:Li-QGM:Antiferro2016arXiv, Bloch:Li-QGM:SpinCharge2016arXiv, Bloch:Rb-QGM:Disorder2016, Zwierlein:K-QGM:ChargeSpinCorr2016arXiv},
otherwise almost impossible to perform.
In the currently developed QGM methods, however, atoms are measured by detecting fluorescent photons from atoms irradiated with near resonant probe light,
resulting in the destruction of the quantum states of atoms such as internal spin states.
In addition, the measurement inevitably induces considerable recoil heating, requiring elaborate cooling scheme in a deep optical lattice.

An ultimate quantum measurement and control such as quantum non-demolition~(QND) measurement
and quantum feedback control is, on the one hand, demonstrated for a single mode of field state with a cavity-quantum-electrodynamics~(QED) system
\cite{Haroche:cavityQED-QND2007,Haroche:QuantumFB2011},
for a collective spin ensemble by a dispersive atom-light interaction
\cite{Jessen:QNDmeasurement2004, Polzik:QNDmeasurement2009, Takano:SpinSqueezing2009, Vuletic:QNDmeasurement2010, Mitchell:QNDmeasurement2012, Inoue:QuantumFB2013, Thompson:QuantumFB2016},
and also for a superconducting quantum bit by a circuit QED system
\cite{IrfanSidiqi:cavityQED-SuperConducting2013}.
In order to realize an ultimate quantum measurement and control for each atom in an optical lattice,
we need to develop a new detection method of QGM which does not rely on the destructive fluorescent measurement.
Promising results along this line were already reported on the detection of
a single atom trapped with a tightly-focused laser beam and a single ion in an ion-trap with a dispersive method
in Ref.~\cite{Kurtsiefer:SingleAtomPhaseShift2009} and Ref.~\cite{Kielpinski:SingleAtomPhaseShift2013}, respectively.
Here we note that, although the use of an optical cavity provides an intriguing sensitivity for a single atom
\cite{Kimble:cavityQEDwithSingleAtom1998, Rempe:CavityQEDwithSingleAtom2004, Vuletic:TrappedSingleAtomInNanoCavity},
this cannot be simply combined with a QGM technique because a cavity spatial mode determines the spatial resolution
and therefore the single-site resolution is not expected.

\begin{figure}[tb]
\centering
\includegraphics{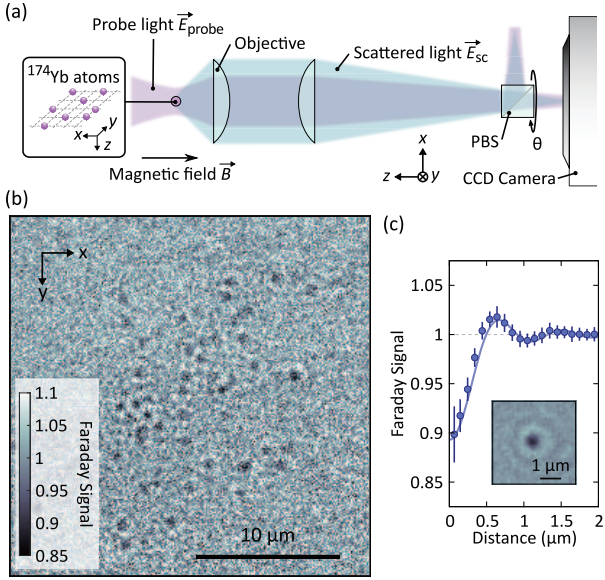}
\caption{(Color online)
			  Faraday imaging.
			  (a) Schematic of our imaging system.
			  We detect a polarization rotation of a  linearly polarized probe light of 399~nm transmitted through $^{174}\Yb$ atoms in a 2D optical lattice
			  with a polarizing beam splitter (PBS) placed in front of a charge-coupled device~(CCD) camera.
			  The high-resolution objective with $\NA=0.75$ is just above the glass cell.
			  The PBS angle $\theta$ is set to be $\pi/4$ for the Faraday imaging.
			  (b) Site-resolved Faraday image of $^{174}\Yb$ atoms.
			  The detuning of the probe beam is $\delta_B=2\pi \times 56\ {\rm MHz}$ and the intensity is $1.3 \times 10^{-2}$ times the saturation intensity,
			  corresponding to the saturation parameter of $0.84 \times 10^{-3}$.
			  The measurement duration is 400~ms.
			  (c) Measured PSF averaged about 30 individual single atoms and azimuthal average of the PSF.
			  The blue line is a fit with Eq.~\eqref{eq:ordiFaraday} with $\NA=0.49(2)$.
}
\label{fig:figure1}
\end{figure}

In this paper, we successfully develop a new detection method of QGM using the dispersive Faraday effect (Faraday QGM),
and  achieve a site-resolved imaging of single isolated atoms in an optical lattice.
The observed Faraday rotation angle reaches 3.0(2)~degrees for a single atom.
We demonstrate the non-destructive feature of this Faraday imaging method by
comparing the detuning dependence of the Faraday signal strength with that of the photon scattering rate.
In addition, we also demonstrate an absorption imaging and a dark field Faraday imaging~(DFFI) of QGM, and reveal the 
different shapes of the point spread functions~(PSFs) for these methods, which are fully explained by theoretical analysis.  
Our result is an important first step towards an ultimate QND measurement and quantum feedback control of a quantum many-body system with a single-site resolution,
which will have significant impacts on quantum information processing and the physics of quantum many-body system \cite{YAshida:QuantumCritical2016arXiv}.

In our experiment we use bosonic ytterbium ($^{174}\Yb$) atoms.
First, we prepare a Bose-Einstein condensate~(BEC) of $^{174}\Yb$ atoms in an optical trap, and then load it into a single layer of two-dimensional~(2D) optical lattice.
In our previous work,
we demonstrate the observation of site-resolved fluorescence imaging of single isolated atoms with a dual molasses technique \cite{RYamamoto:Yb-QGM2016}.
In this work of Faraday QGM, instead of fluorescent photons,
we detect a polarization rotation of a  linearly polarized probe light transmitted through atoms in a 2D optical lattice with a polarizing beam splitter~(PBS) placed in front of a camera,
which is schematically shown in Fig.~\ref{fig:figure1}(a).

A polarization rotation due to the Faraday effect for a single atom can be understood as an interference effect
between a linearly polarized input probe beam $\vec{E}_{\mathrm{probe}}(r)$ and an induced scattering electric field by a single atom.
Based on a diffraction theory \cite{Nano-optics} and a scattering theory \cite{CKurtsiefer:InterfaceSingleAtom2009},
a scattered light field $\vec{E}_{\mathrm{sc}}(r)$ at a detector is described as follows:
\begin{eqnarray}
\vec{E}_{\mathrm{sc}}(r) = \alpha \frac{2J_1(r/\sigma)}{r/\sigma} E_0
\left(
\frac{\hat{e}_+}{1-i(2\delta_B/\varGamma)} + \frac{\hat{e}_-}{1+i(2\delta_B/\varGamma)}
\right), \nonumber \\
\label{eq:Esc}
\end{eqnarray}
where $\delta_B$ represents the detuning from the resonance
(see Supplemental Material S1),
$E_0$ the amplitude of an electric field of an input probe beam,
$\alpha=-\sqrt{3\eta}\NA/2$, $\NA$ the numerical aperture of an objective,
$\eta \equiv [1 - \sqrt{1-\NA^2}(1-\NA^2/4)]/2$ the photon collection efficiency of an objective,
$J_1(x)$ the Bessel function of the first kind,
$\sigma\equiv (k\NA)^{-1}$ the diffraction-limited spatial resolution,
$k$ the wavenumber of probe light,
and $\hat{e}_\pm$ the polarization unit vector for $\sigma_{\pm}$ circularly polarized light, respectively.
Using these expressions,
the total detected field $E_{\mathrm{detect}}(r)$ after a PBS is given as
$E_{\mathrm{detect}}(r) = \left(\vec{E}_{\mathrm{probe}}(r) + \vec{E}_{\mathrm{sc}}(r)\right) \cdot \hat{e}_\theta$,
where $\theta$ and $\hat{e}_\theta$ represent the angle of a PBS with respect to the incident probe polarization and its unit vector, respectively.
Here, in our experimental setup, the beam waist of the probe light is $\sim 37 \ \mathrm{\mu m}$,
much larger than the experimentally measured resolution $\sigma_{\mathrm{exp}}$ of about 120~nm,
enabling us to consider the distribution of probe light as uniform.
Note that the theoretically predicted resolution $\sigma_{\mathrm{ideal}}$ is 85~nm.
A Faraday image, normalized as 1 for the background level, can be described as follows:
\begin{eqnarray}
I_{\mathrm{detect}}(r) &=& \left|E_{\mathrm{detect}}(r) /(E_0\cos\theta) \right|^2
\nonumber \\
&=&
\left|
1+\sqrt{2}\alpha
\dfrac{1 + \left(2\delta_B/  \varGamma \right) \tan\theta}{1+(2 \delta_B / \varGamma)^2}
\dfrac{2J_1(r/\sigma)}{r/\sigma}
\right|^2.
\label{eq:ordiFaraday}
\end{eqnarray}
It is worthwhile to note that this spatial profile of an image of a single atom, namely the PSF, is different from that of the ordinary fluorescence imaging which is given by
\begin{eqnarray}
I_{\mathrm{FL}}(r) \propto
\dfrac{1}{1 + (2\delta_B/\Gamma)^2}
\left(\dfrac{2J_1(r/\sigma)}{r/\sigma}\right)^2.
\label{eq:FluorescenceImaging}
\end{eqnarray}
The difference clearly comes from the presence or absence of the interference
between the probe light $\vec{E}_{\mathrm{probe}}(r)$ and the scattered light $\vec{E}_{\mathrm{sc}}(r)$.
The interference is absent in a fluorescence image.
On the other hand, the interference term $2J_1(x)/x$ is dominant
at a PBS angle $\theta=\pm\pi/4$ for a Faraday image
which is actually observed in our experiment shown in Fig.~\ref{fig:figure1}(c) and discussed below.
We also note that our Faraday imaging method, if applied to an atomic ensemble, is equivalent to
the phase-contrast polarization imaging method developed in Ref.~\cite{Hulet:BECFaraday1997}
and exploited for non-destructive probing of a BEC.

In Fig.~\ref{fig:figure1}(b), we show one illustrative example of the Faraday image
obtained with the measurement setup schematically shown in Fig.~\ref{fig:figure1}(a) with a PBS angle $\theta=\pi/4$.
For easier evaluation of the performance of the Faraday QGM, we intentionally select only several percent of the atoms and prepare a sparse atom cloud for the measurement.
The observed Faraday rotation angle reaches 3.0(2)~degrees for a single atom with the detuning $\delta_B = 2 \pi \times 56\ \mathrm{MHz}$.
Figure~\ref{fig:figure1}(c) shows the measured PSF, obtained by averaging about 30 isolated individual atoms.
We find that our measured PSF is well fitted with the theoretical formula of Eq.~\eqref{eq:ordiFaraday} shown as a blue solid line in Fig.~\ref{fig:figure1}(c).

Here we discuss the current limitation of the Faraday imaging method and its possible solutions.
The Faraday signal is obtained as a result of the interference between the scattered and the probe light beams.
The background level of the Faraday signal is, thus, sensitive to the temporal fluctuation and the spatial inhomogeneity of the intensity and the polarization of the probe beam,
resulting in a relatively poor signal-to-noise ratio. 
This problem can in principle be solved by careful stabilization of the probe beam for its intensity, polarization, and spatial profile.
In the present experiment, to achieve a better signal-to-noise ratio with only intensity stabilization,
the probe beam has a strong intensity which causes the residual heating effect so that cooling during the imaging is required.
An interferometric detection of a weak light using a strong local oscillator for a homodyne detection scheme,
similar to Ref.~\cite{Kurtsiefer:SingleAtomPhaseShift2009},
would enhance the detection sensitivity with a reduced photon scattering rate.
The polarization-squeezed light is also useful for the suppression of the polarization noise below the standard quantum limit
\cite{YTakahashi:QNDmeasurement1999, Kuzmich&Polzik:AtomicContinuousProcessing2003}.
The realization of non-destructive limit of the Faraday QGM would significantly relax the experimental hurdle for a QGM setup,
such as the necessity of an elaborate cooling scheme in an extremely high optical lattice depth during the imaging.
This will even open the possibilities of various atomic species and even molecules for quantum gas microscopy as well as the occupancy-resolved measurement beyond the current parity measurement. 

\begin{figure}[tb]
\centering
\includegraphics{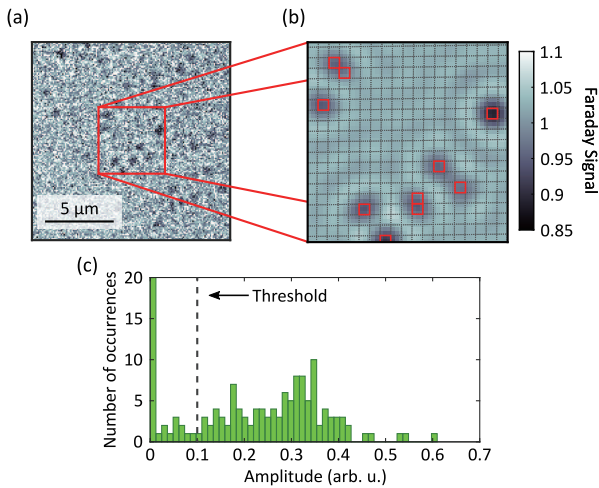}
\caption{(Color online)
			 Deconvolution result of Faraday QGM. 
			 (a) Raw Faraday image of sparsely filled atoms in a lattice.
			 (b) Numerically reconstructed atom distribution on lattice sites.
			 The image is convoluted with the model PSF of Eq.~\eqref{eq:ordiFaraday} and reconstructed atom distribution.
			 Red squares and grey dotted lines represents the atoms and the lattice separations, respectively.
			 (c) Histogram of the fitted amplitudes of the scattered field $\vec{E}_{\mathrm{sc}}(r)$ in each site.
			 A black dashed line shows the threshold of the presence of atoms.
}
\label{fig:Deconvolution}
\end{figure}
We successfully determine the atom distribution by deconvolution of a Faraday image. 
The basic procedure of the deconvolution is almost the same as that in our previous work on fluorescence imaging of QGM, with a PSF of Eq.~\eqref{eq:ordiFaraday} being the main difference.
Figure~\ref{fig:Deconvolution}(a) shows a raw image of Faraday QGM, and in Fig.~\ref{fig:Deconvolution}(b) we show the reconstructed atom distribution convoluted with the model PSF. 
A histogram of the fitted amplitudes of the scattered field $\vec{E}_{\mathrm{sc}}(r)$ in each site is shown in Fig.~\ref{fig:Deconvolution}(c)
and a black dashed line indicates the chosen threshold value.

\begin{figure}[tb]
\centering
\includegraphics{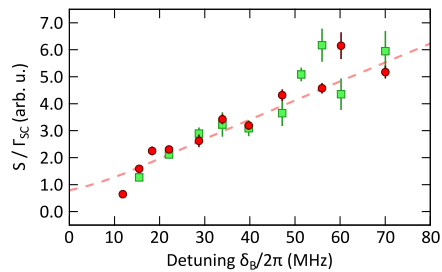}
\caption{(Color online)
			  Ratio of signal strength $S$ to photon scattering rate $\varGamma_{\mathrm{sc}}$ for Faraday imaging.
			  The green squares represent the data obtained from the signals of isolated atoms,
			  and red circles represent the ensemble measurements.
			  A dashed line shows a theoretically fitted curve.
}
\label{fig:RatioHeating}
\end{figure}
The inherent non-destructive nature of the Faraday imaging method originates from the dispersive character of a Faraday effect,
represented by the detuning dependence of the signal expressed by Eq.~\eqref{eq:ordiFaraday}.
The Faraday signal, which is the interference term of Eq.~\eqref{eq:ordiFaraday}, is inversely proportional to the detuning ($\propto 1/\delta_B$) at a large detuning limit.
This should be compared with the destructive effect of photon scattering rate $\varGamma_{\mathrm{sc}}$ by probe light, which is expressed by Eq.~\eqref{eq:FluorescenceImaging}
and proportional to $1/\delta_B^2$ at a large detuning limit.
Therefore, by taking a large detuning, we can improve the ratio of the signal strength to the destructive effect of the photon scattering in Faraday imaging.
In Fig.~\ref{fig:RatioHeating},
we plot the detuning dependence of the ratio of the Faraday imaging signal strength $S$ to the photon scattering rate $\varGamma_{\mathrm{sc}}$ in arbitrary units.
Note that we represent the Faraday imaging signal strength $S$ by the averaged signal of the isolated atoms in the Faraday imaging.
On the one hand, the averaged signal of the isolated atoms in fluorescence imaging taken in the same detuning is used
as a measure of the photon scattering rate $\varGamma_{\mathrm{sc}}$.
The ratios obtained in this way are denoted by green squares.
We also plot the ratios obtained by ensemble measurements as red circles. 
The experimental results are in good agreement with the theoretical prediction represented by a dashed line and in particular show the linear increase with the detuning,
indicating that the Faraday imaging realizes a single-atom observation with a reduced effect of photon scattering.
In fact, the saturation parameter at the detuning of $\delta_B = 2 \pi \times 70\ \mathrm{MHz}$ corresponds to $0.6 \times 10^{-3}$,
almost half of the value of the typical fluorescence imaging.
This is to be contrasted with the case of the fluorescence imaging where the ratio is constant on the detuning, and is not improved.

\begin{figure}[tb]
\centering
\includegraphics{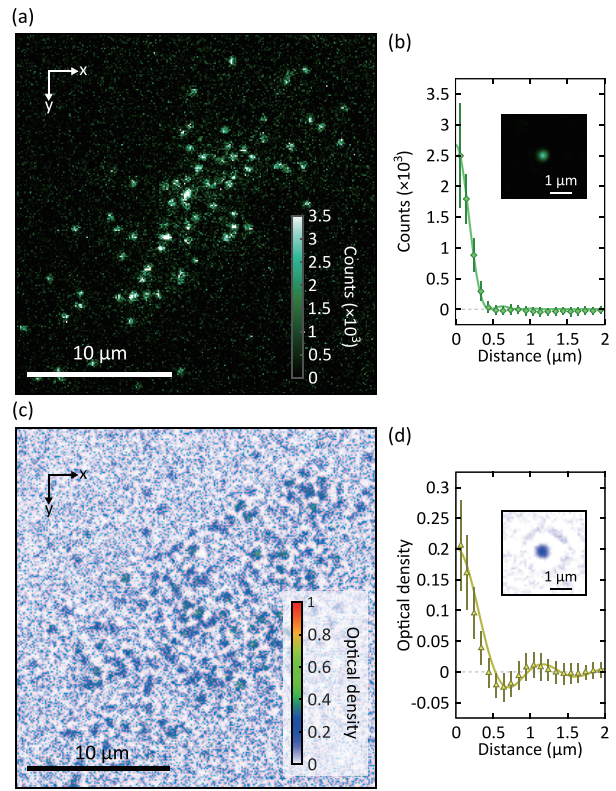}
\caption{(Color online)
			   Site-resolved DFFI and absorption imaging.
			  (a) DFFI ($\theta = \pi/2$).
			  The detuning of the probe beam is $\delta_B=2\pi \times 56\ \mathrm{MHz}$ with the saturation parameter $s_{399} = 1.1 \times 10^{-3}$.
			  (b) Measured PSF averaged about 30 individual single atoms and azimuthal average of DFFI.
			  The green solid line shows a fit with Eq.~\eqref{eq:DFFIPSF}.
			  (c) Absorption imaging ($\theta = 0$).
			  The detuning of the probe beam is $\delta_B = 2\pi \times 11\ \mathrm{MHz}$ with the saturation parameter $s_{399} = 2.9 \times 10^{-3}$.
			  (e) Measured PSF averaged about 60 individual single atoms and azimuthal average of absorption image.
			  The yellow solid line shows a fit with $-\log\left[I_{\mathrm{detect}}(r)\right]$
			  where $I_{\mathrm{detect}}(r)$ is given by Eq.~\eqref{eq:ordiFaraday}.
			  Each measurement takes the duration of 400~ms.
}
\label{fig:figure4}
\end{figure}

In addition to the Faraday imaging with the PBS angle $\theta=\pi/4$,
we also demonstrate a different type of Faraday imaging of DFFI \cite{Jacob:darkFaraday2013} by setting $\theta=\pi/2$ in the setup of Fig.~\ref{fig:figure1}(a).  
In this case, all of the probe light is reflected by the PBS and only the scattered light can be transmitted through and detected at a camera.
This configuration of DFFI enables us to obtain a back-ground-free signal like a fluorescence signal.
Again, for easier evaluation,
only several percent of the atoms are selected and cooling beams are applied to suppress the heating effect and to keep atoms within the respective lattice sites.
Figure~\ref{fig:figure4}(a) shows the DFFI signal of site-resolved image of single atoms.
Here the detuning is $2\pi \times 56\ \mathrm{MHz}$, which is the same as Fig.~\ref{fig:figure1}(b) of the Faraday imaging.
Although this DFFI signal looks quite similar to that of fluorescence imaging, the DFFI signal originates from a dispersive interaction just like a Faraday signal.
Figure~\ref{fig:figure4}(b) shows the measured PSF, obtained by averaging about 30 individual atoms.
We find that the measured PSF is well fitted with the theoretical formula
given by
\begin{equation}
I_{\mathrm{DFFI}}(r) \propto
\left(
\dfrac{2 \delta_B/\varGamma}{1 + (2\delta_B/\varGamma)^2}
\dfrac{2J_1(r/\sigma)}{r/\sigma}
\right)^2,
\label{eq:DFFIPSF}
\end{equation}
and a green solid line in Fig.~\ref{fig:figure4}(b) shows a fit with Eq.~\eqref{eq:DFFIPSF}.
We note that the DFFI signal has a detuning dependence of $\propto 1/\delta_B^2$ at a large detuning limit, and has no non-destructive nature.
The experimental results in fact show the saturation of the ratio of the DFFI signal to the photon scattering rate $\varGamma_{\mathrm{sc}}$ at larger detunings,
consistent with the theoretical prediction, indicating that the DFFI has no merit to realize a single-atom observation with a reduced effect of photon scattering.

Moreover,
we demonstrate an absorption imaging by setting the PBS angle $\theta=0$, which is the standard setup for an ensemble measurement.  
In this case, similarly to the Faraday imaging, a probe light makes destructive (and also constructive) interference with scattered light.
Figure~\ref{fig:figure4}(c) is the absorption imaging signal, which clearly shows a site-resolved image of single atoms.
Here the detuning is taken as $2\pi \times 11\ \mathrm{MHz}$ within the linewidth of the probe transition.
Figure~\ref{fig:figure4}(d) shows the measured PSF, obtained by averaging about 60 individual atoms,
which reveals the interference feature like the Faraday imaging.
Again we find that our measured PSF is well fitted with the theoretical formula of $-\log\left[I_{\mathrm{detect}}(r)\right]$
shown as a yellow solid line in Fig.~\ref{fig:figure4}(d) and
a peak optical density of the PSF reaches 0.20(2) corresponding to a maximum extinction of 18(1)\% by a single atom.
This value is much larger than that of the previous works for single atoms and ions \cite{Kurtsiefer:SingleAtomDetection2008, Kielpinski:AIofSingleAtom2012}.

In conclusion, we successfully demonstrate site-resolved imaging of single atoms with the Faraday effect.
The observed Faraday rotation angle reaches 3.0(2)~degrees for a single atom.
We demonstrate the non-destructive feature of this Faraday imaging method by investigating the detuning dependence of the signal.
In addition, we demonstrate absorption imaging and DFFI of QGM,
and reveal the different shapes of PSFs for these imaging methods,
which are fully explained by theoretical analysis.
Our result is an important step towards an ultimate QND measurement with a single-site resolution.
It will furthermore open up the possibilities for quantum feedback control of individual atoms in a quantum many-body system
which will have significant impact on quantum information processing and the physics of quantum many-body system.

The authors are grateful to E. Chae and S. Yamanaka for careful reading of the manuscript.
This work was supported by the Grant-in-Aid for Scientific Research of JSPS
(Nos. 13J00122, 25220711, 26247064, 26610121, 16H00990, 16H01053)
and the Impulsing Paradigm Change through Disruptive Technologies (ImPACT) program.

%

\clearpage
\onecolumngrid
\section{Supplemental Material}
\setcounter{figure}{0}
\setcounter{equation}{0}
\renewcommand{\thefigure}{S\arabic{figure}}
\renewcommand{\theequation}{S\arabic{equation}}

\subsection{Supplemental Material S1: Low-lying energy levels of ytterbium atom}
\begin{figure}[htb]
\centering
\includegraphics{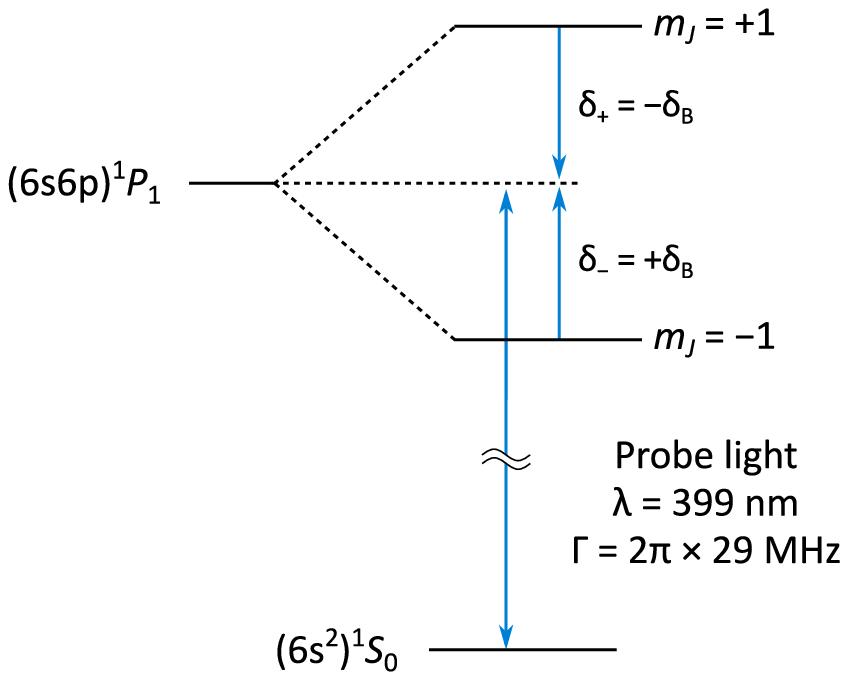}
\caption{(Color online)
			  Low-lying energy levels of $^{174}\mathrm{Yb}$ atoms relevant for probing.
			  The frequency of the probe beam is set at the center of the ${}^{1}S_0\mhyph{}^{1}P_1 (m_J=1)$ and ${}^{1}S_0\mhyph{}^{1}P_1 (m_J=-1)$ transitions.
}
\label{fig:figureS1}
\end{figure}

Low-lying energy levels associated with probing are shown in Fig.~\ref{fig:figureS1}.
We apply a magnetic field $\vec{B}$ for inducing a Faraday effect and $\vec{B}$ is almost parallel to the $z$-axis which is the propagation direction of a probe beam
(See Fig.~1(a) of the main text).
A linearly polarized probe beam is near resonant with the ${}^{1}S_0\mhyph{}^{1}P_1 (m_J=\pm1)$ transition (transition wavelength $\lambda = 399\ \mathrm{nm}$,
natural linewidth $\varGamma =2\pi\times 29\ \mathrm{MHz}$).
For most of the measurements in this work we set the frequency of  a probe beam at the center of the ${}^{1}S_0\mhyph{}^{1}P_1 (m_J=\pm1)$ transitions, otherwise noticed.
Thus, the detuning of a probe laser beam with respect to the ${}^{1}S_0\mhyph{}^{1}P_1 (m_J=\pm1)$ transition is $\mp \delta_B$, respectively.
Here $\delta_B= g_J \mu_\mathrm{B} |\vec{B}| / \hbar$ is a Zeeman shift in the ${}^1P_1(m_J=\pm1)$ state due to the magnetic field $\vec{B}$,
$g_J$ the Land\`{e} g-factor of $^1P_1$ state, and $\mu_B$ a Bohr magneton (Fig.~\ref{fig:figureS1}).
Since the applied magnetic field is almost parallel to the $z$-axis, we have negligible excitation for the  ${}^{1}S_0\mhyph{}^{1}P_1 (m_J=0)$ transition.

\subsection{Supplemental Material S2: Optical Spectra of Faraday imaging, DFFI, and absorption imaging}
\begin{figure}[htb]
\centering
\includegraphics[scale=1.2]{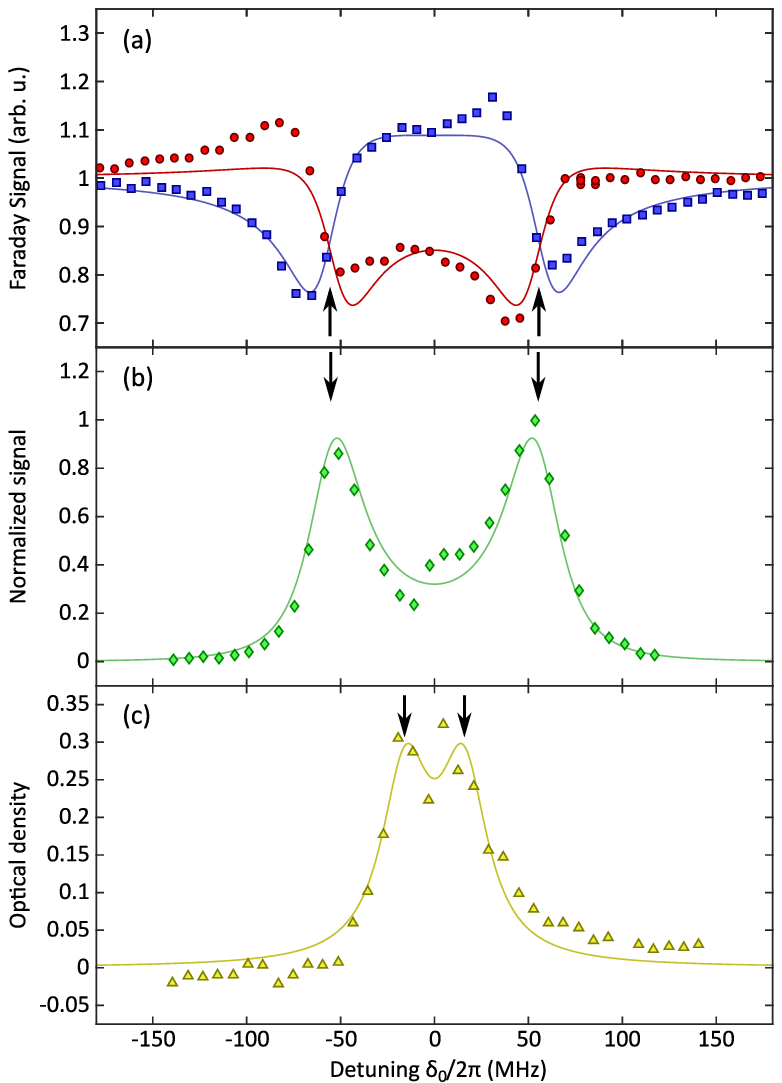}
\caption{(Color online)
			  Optical spectra of (a) Faraday imaging ($\theta = \pm \pi/4$), (b) DFFI ($\theta = \pi/2$), and (c) absorption imaging ($\theta = 0$).
			  A magnetic field of a 40~G is applied for Faraday imaging and DFFI, and 8~G for absorption imaging.
              The resonance positions are indicated by arrows in the figure.
			  (a) A red (blue) curve shows the spectrum with a PBS angle $\theta = \pi/4\ (-\pi/4)$.
			  (b) The solid line shows a fit with Eq.~\eqref{eq:DFFIspectrum}.
			  The signal strength in each spectrum corresponds to the total counts of the image.
}
\label{fig:Spectra}
\end{figure}

Figure~\ref{fig:Spectra} shows optical spectra of (a)~Faraday imaging, (b)~DFFI, and (c)~absorption imaging, respectively,
where the total counts in the respective image is plotted as a function of an applied probe laser frequency. 
Here a magnetic field of 40~G is applied for the Faraday imaging and the DFFI, and 8~G for the absorption imaging.
The resonance positions are indicated by arrows in the figure.

The Faraday imaging shows a dispersive frequency dependence around the ${}^{1}S_0\mhyph{}^{1}P_1 (m_J=\pm1)$ resonances ($\delta_0 = \pm 2\pi \times 58\ \mathrm{MHz}$),
which can be fitted with a following equation:
\begin{equation}
A_{\mathrm{FI}}(\delta_0) = \left|
\dfrac{\left(\vec{E}_{\mathrm{probe}} + \vec{E}_{\mathrm{sc}}(\delta_0) \right) \cdot \hat{e}_\theta}
{\vec{E}_{\mathrm{probe}} \cdot \hat{e}_\theta}
\right|^2,
\end{equation}
where
\begin{equation}
\vec{E}_{\mathrm{sc}}(\delta_0)
\propto
\left(
\dfrac{\hat{e}_+}{1+i2(\delta_0 - \delta_B)/\varGamma}
+\dfrac{\hat{e}_-}{1+i2(\delta_0 + \delta_B)/\varGamma}
\right).
\end{equation}
In Fig.~\ref{fig:Spectra}(a), a red curve shows a fit with $\theta = \pi/4$ and a blue one a fit with $\theta = -\pi/4$.

The signal of DFFI can be described by 
\begin{equation}
A_{\mathrm{DFFI}}(\delta_0) \propto \left| \vec{E}_{\mathrm{sc}}(\delta_0) \cdot \hat{e}_{\pi/2} \right|^2
\propto \left|
\dfrac{1}{1+i2(\delta_0 - \delta_B)/\varGamma}
-\dfrac{1}{1+i2(\delta_0 + \delta_B)/\varGamma}
\right|^2.
\label{eq:DFFIspectrum}
\end{equation}
The solid line in Fig.~\ref{fig:Spectra}(b) shows a fit with Eq.~\eqref{eq:DFFIspectrum}.

The absorption imaging shows a resonant character,
which can be fitted with $-\log\left[A_{\mathrm{FI}}(\delta_0)\right]$ with $\theta = 0$ shown in Fig.~\ref{fig:Spectra}(c).

\subsection{Supplemental Material S3: Faraday rotation angle of a single atom}
\begin{figure}[htb]
\centering
\includegraphics[scale=1.5]{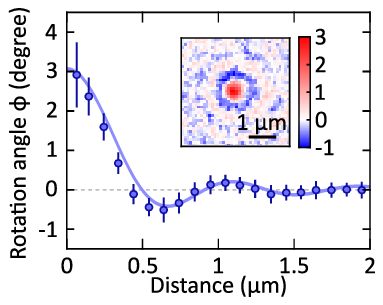}
\caption{(Color online)
			Azimuthal average of Faraday rotation angle evaluated with Eq.~\eqref{eq:FaradayAangle}.
			The detuning $\delta_B$ of the probe light is $2\pi \times 56\ \mathrm{MHz}$ with the saturation parameter $s_{399} = 0.84 \times 10^{-3}$.
			A peak Faraday rotation angle reaches 3.0(2)~degrees.
			}
\label{fig:PSFangle}
\end{figure}

$I_{\mathrm{detect}}(r)$ given in Eq.~(2) is also described as $I_{\mathrm{detect}}(r) = \left[\cos(\theta + \phi(r))/\cos\theta\right]^2$,
by introducing a position-dependent Faraday rotation angle
$\phi(r)$ defined as $\vec{E}_{\mathrm{probe}}(r)+\vec{E}_{\mathrm{sc}}(r)=E_{0}(e^{+i\phi(r)} \hat{e}_{+} + e^{-i\phi(r)} \hat{e}_{-})/\sqrt{2}$.
Therefore, $\phi(r)$ can be calculated with a following equation:
\begin{equation}
\phi(r) = \cos^{-1}\left[\sqrt{I_{\mathrm{detect}}(r) \cos^2\theta}\right] - \theta.
\label{eq:FaradayAangle}
\end{equation}
From the data of Fig.~1(c) and Eq.~\eqref{eq:FaradayAangle} with $\theta = \pi/4$,
we evaluate the spatial distribution of the Faraday rotation angle of a single atom and its azimutial average, as shown in Fig.~\ref{fig:PSFangle}.

\subsection{Supplemental Material S4: Effect of a probe beam for Faraday QGM}
\begin{figure}[htb]
\centering
\includegraphics{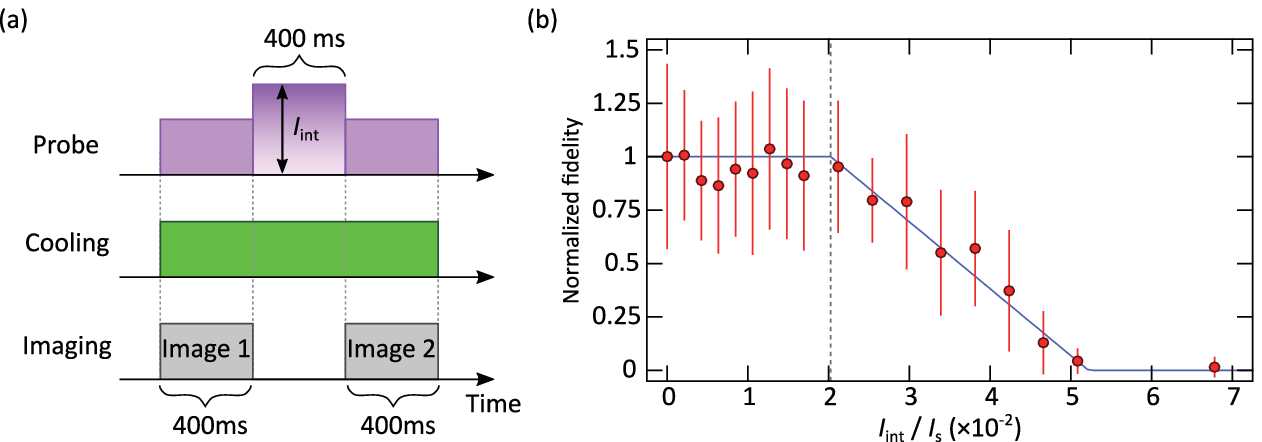}
\caption{(Color online)
			Effect of the probe light for Faraday QGM.
			(a) Timing of taking two consecutive images and varying the probe intensity.
			The duration of the exposure and the interval time is 400~ms.
			(b) Measured fidelity normalized by the fidelity
			without probe light is plotted for various probe intensities.
			Below the intensity $I_{\mathrm{int}} = 2\times 10^{-2} I_{\mathrm{s}}$,
			the normalized fidelity takes almost one.
			Note that $I_{\mathrm{s}}$ represents the saturation intensity of the probe beam.
			}
\label{fig:HeatingEffect}
\end{figure}

Usually the fidelity of the imaging can be evaluated by taking two successive images of the same atoms and comparing the atom distributions.
The fidelity deduced from such a method includes the fidelity of the deconvolution procedure, which will make a large contribution
in the current Faraday imaging, especially at low probe intensities.
Here, in order to purely extract the effect of the probe light for Faraday imaging,
we apply a probe pulse with the same detuning as the Faraday imaging and with varying intensities during the  400~ms interval between the two images.
The timing of taking two consecutive images and applying the probe beam is shown schematically in Fig.~\ref{fig:HeatingEffect}(a).
The consecutive two images to determine the atom distributions are taken by setting the PBS angle to $\pi/2$ (DFFI)
so that we can get the background-free image similar to the fluorescence images.
Note that the cooling light is also applied to suppress the residual heating effect as in the Faraday imaging.
In Fig.~\ref{fig:HeatingEffect}(b), we show the fidelity normalized by that without the probe light during two images.
We find almost no change of pinned, loss, and hopping fractions when the probe intensity is below $2 \times 10^{-2}$ times the saturation intensity.
We note that most of the measurements in this paper are done in this regime.
Above this intensity, we find almost linear increase of the loss and hopping fractions.
This behavior is reasonable when considered in terms of the saturation parameter.
The observed critical value corresponds to the saturation parameter of $s_{399} \sim 10^{-3}$ which is consistent with that observed in the previous experiment,
where the cooling is balanced with the heating effect of the probe beam.

\end{document}